\documentclass[iop]{emulateapj}
\usepackage{epsfig}
\usepackage{apjfonts}
\usepackage{aas_macros}
\usepackage{amsmath}

\begin{document}

\title{Magnetic-distortion-induced ellipticity and gravitational wave radiation of neutron stars: millisecond magnetars in short GRBs, Galactic pulsars, and magnetars}
\author{He Gao$^{1,*}$, Zhoujian Cao$^{1}$, and Bing Zhang$^{2,3,4}$}
\affiliation{
$^1$Department of Astronomy, Beijing Normal University, Beijing 100875, China; gaohe@bnu.edu.cn
\\$^2$Department of Physics and Astronomy, University of Nevada Las Vegas, NV 89154, USA;\\
  $^3$Department of Astronomy, School of Physics, Peking University, Beijing 100871, China; \\
  $^4$Kavli Institute of Astronomy and Astrophysics, Peking University, Beijing 100871, China.}

\begin{abstract}
Neutron stars may sustain a non-axisymmetric deformation due to magnetic distortion and are potential sources of continuous gravitational waves (GWs) for ground-based interferometric detectors. With decades of searches using available GW detectors,
no evidence of a GW signal from any pulsar has been observed.  Progressively stringent upper limits of ellipticity have been placed on Galactic pulsars.  In this work, we use the ellipticity inferred from the putative millisecond magnetars in short gamma-ray bursts (SGRBs) to estimate their detectability by current and future GW detectors. For $\sim 1$ ms magnetars inferred from the SGRB data, the detection horizon is $\sim 30$ Mpc and $\sim 600$ Mpc for advanced LIGO (aLIGO) and Einstein Telescope (ET), respectively. Using the ellipticity of SGRB millisecond magnetars as calibration, we estimate the ellipticity and gravitational wave strain of Galactic pulsars and magnetars assuming that the ellipticity is magnetic-distortion-induced. We find that the results are consistent with the null detection results of Galactic pulsars and magnetars with the aLIGO O1. We further predict that the GW signals from these pulsars/magnetars may not be detectable by the currently designed aLIGO detector. The ET detector may be able to detect some relatively low frequency signals ($<50$ Hz) from some of these pulsars. Limited by its design sensitivity, the eLISA detector seems not suitable for detecting the signals from Galactic pulsars and magnetars. 
\end{abstract}


\section {Introduction}

The Laser Interferometer Gravitational-wave Observatory (LIGO) team have announced two direct detections of gravitational wave (GW) events (GW 150914 and GW 151226) from binary black hole mergers \citep{abbott16a,abbott16b}. 
This marked the beginning of the GW astronomy. Besides the primary targets of inspiral and mergers of compact object binary (NS-NS, NS-BH, BH-BH) systems, another potential target of continuous GW emission for the ground-based GW detectors, such as Advanced LIGO \citep{abbott09}, Advanced VIRGO \citep{acernese08} and KAGRA \citep{kuroda10} interferometers, are rapidly rotating neutron stars, as long as the NSs sustain a significant non-axisymmetric deformation \cite[][and reference therein]{aasi14}. 

Several mechanisms to induce NS asymmetries have been suggested in the literature. First, the crust of a NS is solid and elastic. The shape of the crust depends on many factors, such as the original formation history and accretion history of the NS, star quakes, and the equation of state of NS \citep{ushomirsky00,haskell06}. The deformation of the crust would not be easily smoothed under the effect of rotation, since it could be supported by anisotropic stresses in the solid. Secondly, gravitational radiation reaction \cite[][for a review]{owen02,andersson03} or nuclear matter viscosity \cite[][and references therein]{bonazzola96a} may drive non-axisymmetric instabilities in rapidly rotating neutron stars, which could also produce asymmetries in neutron stars. Finally, neutron stars are known to have relatively large magnetic fields, and the anisotropy of the magnetic pressure would also distort the star. When the magnetic axis is not aligned with the rotation axis, the deformation would not be axisymmetric \citep{ostriker69,bonazzola96b,konno00,ioka04,tomimura05,stella05,haskell08,dall09,mastrano11}. This last mechanism likely plays an important role in defining $\epsilon$ in various astrophysical contexts, especially in magnetars.

Quadrupole deformation of the neutron star is characterized by the ellipticity $\epsilon$, which is defined by 
\begin{eqnarray}
\epsilon=\frac{\rm equatorial~radius-polar~radius}{\rm mean~radius}
\end{eqnarray}
A neutron star with rotation period $P$ and ellipticity $\epsilon$ radiates gravitational waves at a frequency $f=2/P$ with an energy loss rate\footnote{In general, gravitational radiation is emitted at the spin frequency and its octave, and the total energy loss rate is $\dot{E}_{\rm GW}=-\frac{2}{5}\frac{GI^2\epsilon^2\Omega^6}{c^5}\sin^2\chi(1+15\sin^2\chi)$, where $\chi$ is the tilt angle between the spin axis and the non-symmetric axis \cite[][and reference therein]{cutler01}. For simplicity, in this work we adopt an orthogonal rotator ($\chi=\pi/2$), and assume that GWs only emit at twice of the spin frequency.} \citep{shapiro83,usov92,zhang01}
\begin{eqnarray}
\dot{E}_{\rm GW}=-\frac{32GI^2\epsilon^2\Omega^6}{5c^5},
\label{dotE}
\end{eqnarray}
where $\Omega=2\pi /P$ is the angular frequency, $I$ is the moment of inertia of the NS.

In principle, the value of $\epsilon$ may be measured from observations once the GW radiation from one particular NS is detected. Decades of searches with various GW detectors (e.g. initial LIGO, Virgo, GEO 600, and the first observing run of the Advanced LIGO (aLIGO) detectors) targeted on a selection of known Galactic pulsars \cite[][and references therein]{aasi14,abbott17}, however, did not detect any GW signals due to the limitation of sensitivity of the current detectors. 
More and more stringent upper limits on the value of $\epsilon$ have been set for these pulsars. For eight pulsars, the resulted upper limits already surpass their spin-down limits (which attributing all the spin-down luminosity to GW radiation lost) \citep{abbott17}. 

Recently, an indirect method has been proposed to estimate the $\epsilon$ value for a particular class of NSs, i.e., the rapidly spinning, strongly magnetized, supramassive neutron stars (henceforth, millisecond magnetar). A millisecond magnetar has long been proposed to be a possible central engine for gamma-ray bursts (GRBs) \citep{usov92,dai98,zhang01,dai06,gao06,metzger11} and electromagnetic counterparts of NS-NS mergers \citep{zhang13,gao13,yu13,metzger14}. The model is especially relevant to some short GRBs with soft $\gamma$-ray extended emission \citep{norris06,sakamoto11} or GRBs with an internal X-ray plateau followed by a very rapid decay \citep{troja07,rowlinson10,lv15}. These features mark the abrupt cessation of the central engine, likely due to the collapse of a supramassive NS into a BH \citep{rowlinson10,lv14,lv15}. When modeling X-ray plateau of short GRBs, \cite{fan13} noticed that the observed duration of internal X-ray plateau is shorter than that expected in the magnetic dipole radiation scenario. They then suggested that GW radiation likely dominates the loss of rotational energy for these millisecond magnetars. By investigating some particular cases of short GRBs, they suggested that the ellipticity and dipole magnetic field strength ($B_{\rm dip}$) for the supramassive magnetars are around 0.01 and $10^{15}$ G, respectively. They also claimed that the GWs from such sources may be detectable with the proposed Einstein Telescope \citep{punturo10}. Later, \cite{gao16} used the statistical observational properties of Swift SGRBs and the mass distribution of Galactic double neutron star systems to systematically place constraints on the neutron star equation of state (EoS) and the properties of the post-merger product. They found that when the SGRB central engine is a supramassive NS, in order to reproduce the distributions of internal X-ray plateau luminosity and break time, the ellipticity of the millisecond magnetar need to be in the range of $0.004-0.007$, and the dipole magnetic field strength of the NS is typically $10^{15}$ G. Significant GW radiation is expected to be released after the merger. This conclusion applies to a range of EoSs \citep{li16}. \cite{lasky16} explored the physically motivated $\epsilon$ via the spin-flip mechanism. Even though the relatively large value $\epsilon \sim 0.01$ inferred by \cite{fan13} may not be physically unattainable, the value ($\epsilon \sim 0.004-0.007$) inferred by \cite{gao16} is marginally consistent with the range of $\epsilon$ suggested by \cite{lasky16}.

This method of inferring $\epsilon$ is based on the electromagnetic observations of SGRB X-ray afterglows. It is of great interest to investigate the consistency between this result with the GW observations of Galactic pulsars and magnetars. This is the purpose of this paper.

\section{General formalism}

Among various mechanisms, magnetic distortion likely plays a dominant role to maintain a relatively large $\epsilon$ for a millisecond magnetar. We focus on this possibility. According to previous analytical and numerical studies, the magnetic distortion of a NS depends on the strength and the configuration of the magnetic fields (including the inclination angle and the toroidal-to-poloidal ratio) \citep{bonazzola96b,haskell08}. In general, one may parameterize that
\begin{eqnarray}
\epsilon =\beta \bar{B}^2,
\end{eqnarray}
where 
\begin{eqnarray}
\bar{B}^2=\frac{1}{V}\int \textbf{B}^2dV
\end{eqnarray}
scales with the volume average of magnetic pressure, $V$ is the volume of the star, and the coefficient $\beta$ contains the information of the magnetic field configurations. For simplicity, we connect $\bar{B}$ and $B_{\rm dip}$ by defining
\begin{eqnarray}
B_{\rm dip}=\eta\bar{B},
\end{eqnarray}
with $0\leq\eta\leq1$, where $\eta=0$ and 1 represent a star with a purely toroidal and poloidal field component, respectively.

Given the dipole magnetic field strength $B_{\rm dip,m}$ and the ellipticity $\epsilon_m$ for the millisecond magnetars, the ellipticity for a Galactic pulsar ($\epsilon_p$) with dipole magnetic field strength $B_{\rm dip,p}$ may be estimated as (assuming that magnetic distortion is the dominant mechanism to define their respective $\epsilon$)
\begin{eqnarray}
\epsilon_p=5\times10^{-9}\frac{\epsilon_m}{0.005}\left(\frac{\eta_m}{\eta_p}\right)^{2}\left(\frac{B_{\rm dip,p,12}}{B_{\rm dip,m,15}}\right)^{2},
\label{eq:ep}
\end{eqnarray}
where $\eta_p$ and $\eta_m$ represent the configuration of the magnetic fields for the pulsar and the millisecond magnetar, respectively. 

It is worth pointing out that the value of $\epsilon_p$ must not be larger than the spin-down limit ($\epsilon_{\rm sd}$), which is determined by equating the power radiated through gravitational waves to the observed spin-down luminosity of the pulsar, i.e. $\dot{E}_{\rm sd}=I\Omega\dot{\Omega}=\dot{E}_{\rm GW}$. This gives
\begin{eqnarray}
\epsilon_{\rm sd}&=&\sqrt{\left|\frac{5c^5\dot{\Omega}}{32GI\Omega^5}\right|}  \nonumber  \\
&=&1.9\times10^{-7}\left(\frac{P}{1~\rm {ms}}\right)^{3/2}\left(\frac{\dot{P}}{10^{-15}}\right)^{1/2},
\label{eq:ep2}
\end{eqnarray}
where $P$ and $\dot{P}$ are the period and period derivative of the pulsar, respectively.

The characteristic gravitational wave amplitude of a rotating magnetized NS with ellipticity $\epsilon$ and rotation frequency $\Omega$ can be estimated as \citep{corsi09}
\begin{eqnarray}
h_c=fh(t)\sqrt{\frac{dt}{df}},
\label{eq:hc}
\end{eqnarray}
where $f=\Omega/\pi$,
\begin{eqnarray}
h(t)=\frac{4G\Omega^2}{c^4d}I\epsilon,
\label{eq:ht}
\end{eqnarray}
and $d$ is the distance to the source. 

For a given pulsar, once its characteristic gravitational wave amplitude $h_c$ is detected, the $\epsilon$ value could be directly measured. On the other hand, if no GW signal is detected, an upper limit on $\epsilon$ can be set by comparing $h_c$ with the noise level of the GW detector
\begin{eqnarray}
h_{\rm rms}=\left[fS_n(f)\right]^{1/2},
\label{eq:hrms}
\end{eqnarray}
where $S_n(f)$ is the power spectral density (PSD) of the detector noise. We consider aLIGO, ET and eLISA \citep{amaro12} detectors for a single detector analysis. The PSD for aLIGO O1 and the designed PSD for eLISA are adopted from the respective websites of these collaborations\footnote{For aLIGO, see \url{https: //dcc.ligo.org/LIGO- G1600150/public}; and for eLISA, see \url{https://www.elisascience.org/articles/elisa-mission/lisa-white-paper.}}. For the designed PSD of aLIGO, we adopt the following analytical model \citep{arun05,sun15}
\begin{eqnarray}
S_n(f)=S_0\left[x^{-4.14}-5x^{-2}+\frac{111(1-x^2+x^4/2)}{1+x^2/2}\right]
\label{eq:ligo}
\end{eqnarray} 
for $f\geq20$ Hz, where $x=f/f_0$, $f_0=215$ Hz, and $S_0=10^{-49}~\rm Hz^{-1}$.  When $f<20$ Hz, $S_n(f)=\infty$ is adopted. 

For the designed PSD of ET, we adopt the following analytical model \citep{mishra10,sun15}
\begin{eqnarray}
\begin{split}
S_n(f)&=S_0\biggl[ 2.39\times10^{-27}x^{-15.67}+0.349x^{-2.145}\\
&~~~+1.76x^{-0.12}+0.409x^{1.1}\biggr]^{2} 
\label{eq:ET}
\end{split}
\end{eqnarray} 
for $f\geq10$ Hz, where $x=f/f_0$, $f_0=100$ Hz, and $S_0=10^{-50}~\rm Hz^{-1}$.  When $f<10$ Hz, $S_n(f)=\infty$ is adopted. 

\section{Results}

\subsection{Millisecond magnetars in SGRBs}

Based on the results from \cite{gao16}, we adopt $B_{\rm dip,m}=10^{15}$ G and $\epsilon_m=0.005$ as the dipole magnetic field strength and the ellipticity for the millisecond magnetars. 
With Eqs. \ref{eq:hc} and \ref{eq:ht}, one can estimate the characteristic gravitational wave amplitude $h_c$ for millisecond magnetars in SGRBs. Comparing the value of $h_c$ with the noise level of the GW detectors $h_{\rm rms}$, one can estimate the detection horizon of GW signals from these millisecond magnetars, i.e.
\begin{eqnarray}
d &\leq & \left(\frac{5IG}{Pc^{3}}\right)^{1/2}h_{\rm rms}^{-1}\\ \nonumber
&\lesssim& 360 ~{\rm Mpc}~\left(\frac{h_{\rm rms}}{10^{-22}}\right)^{-1}\left(\frac{I}{10^{45}~{\rm g~cm^{2}}}\right)^{1/2}\left(\frac{P}{1~{\rm ms}}\right)^{-1/2}
\end{eqnarray}

\begin{figure}
\centering
\includegraphics[width=0.45\textwidth]{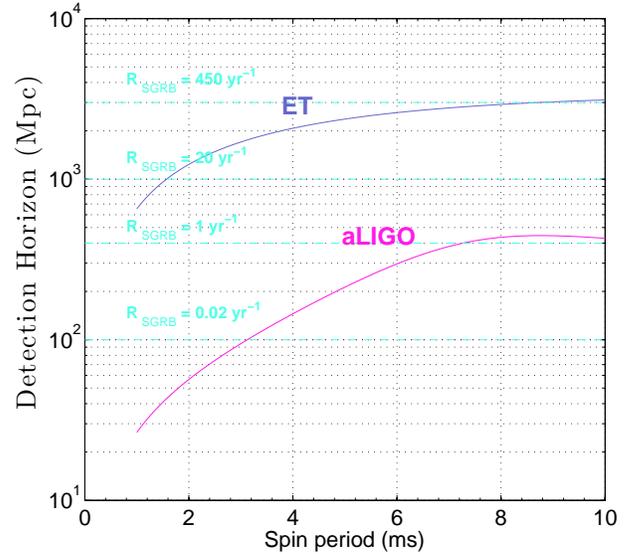}
\caption{Detection horizon of GW signals from SGRB central millisecond magnetars for aLIGO and ET. For reference, we plot the SGRB detection rate for an all-sky gamma-ray monitor, which is extrapolated with the ``local" SGRB detection rate of $ \sim 4 \rm~Gpc^{-3}~yr^{-1}$  above $10^{50} {\rm erg~s^{-1}}$ \citep{wanderman15,sun15b}.} 
\label{fig:3}
\end{figure}

Substituting Eqs. \ref{eq:ligo} and \ref{eq:ET}, we plot the detection horizon of GW signals from millisecond magnetars for aLIGO and ET  (Figure \ref{fig:3}). We can see that the aLIGO horizon for such a signal could be up to 400 Mpc, while the ET horizon could be up to 3 Gpc, both for relatively-slowly-spinning magnetars ($p\geq8~\rm ms$). For $\sim1$ ms magnetars as inferred from the SGRB data \citep{gao16}, the detection horizons for aLIGO and ET are $\sim 30$ Mpc and $\sim 600$ Mpc, respectively. The corresponding SGRB detection rate \citep{wanderman15,sun15b} is low for aLIGO, but is reasonably high for ET (Fig. \ref{fig:3}).

\subsection{Galactic pulsars and magnetars}

With the calibration from millisecond magnetars inferred from SGRB data \citep{gao16}, the ellipticity for a pulsar ($\epsilon_p$) with dipole magnetic field strength $B_{\rm dip,p}$ could be extrapolated from Eq. \ref{eq:ep}. 

\begin{figure}
\centering
\includegraphics[width=0.45\textwidth]{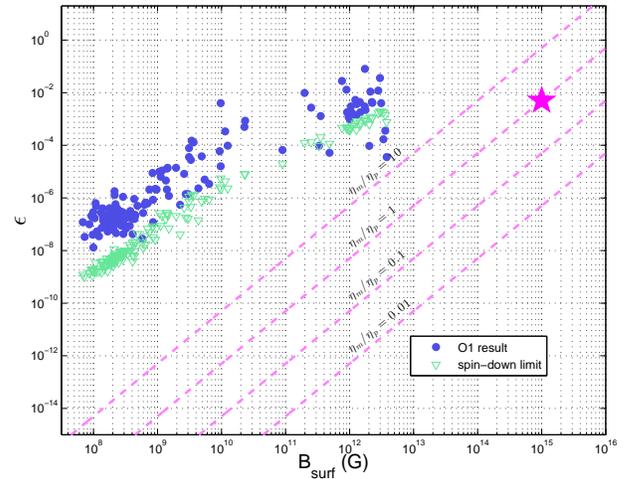}
\caption{A comparison between the pulsar ellipticity inferred from the SGRB data extrapolation and the upper limits placed from the aLIGO O1 non-detections. The pink star marks the millisecond magnetar ellipticity inferred from the SGRB data, and the pink dashed lines represent its extrapolated results for different $\eta_m/\eta_p$ values. The blue dots remark the aLIGO O1 upper limits for Galactic pulsars and the green triangles present their spin-down upper limits.}
\label{fig:1}
\end{figure}

In Figure \ref{fig:1}, we plot the extrapolated results for different $\eta_m/\eta_p$ values. Comparing with the aLIGO O1 results, we find that the millisecond magnetar ellipticity value inferred from the SGRB data would be consistent with the aLIGO O1 results, as long as $\eta_m$ is not larger than $\eta_p$ by more than one order of magnitude. Since the toroidal field is more important (smaller $\eta$) in rapid rotators, it is essentially impossible to have $\eta_m/\eta_p >1$. Our results therefore suggest that the non-detection of Galactic pulsars by aLIGO O1 is naturally expected given the $\epsilon_m$ inferred from the SGRB data. 

With the calibration from Eq. \ref{eq:ep}, we can estimate the characteristic gravitational wave amplitude $h_c$ for known pulsars and investigate their detectability with the current and future GW detectors.  We estimate the expected
$h_c$ values for all known pulsars listed in v1.56 of the ATNF pulsar catalogue \citep{manchester05}\footnote{\url{http://www.atnf.csiro.au/research/pulsar/psrcat/}} and all known magnetars listed in the McGILL magnetar catalogue \citep{olausen14}\footnote{\url{http://www.physics.mcgill.ca/~pulsar/magnetar/main.html}}. 
For each pulsar, we apply Eq. \ref{eq:ep} to estimate its ellipticity (and make sure that it does not exceed Eq. \ref{eq:ep2}), and then apply Equations \ref{eq:hc} and \ref{eq:ht} to calculate its characteristic gravitational wave amplitude $h_c$. In the estimation, $B_{\rm dip}$,  $P$,  $\dot{P}$ and the distance $d$ of each pulsar is used.

\begin{figure}
\centering
\includegraphics[width=0.45\textwidth]{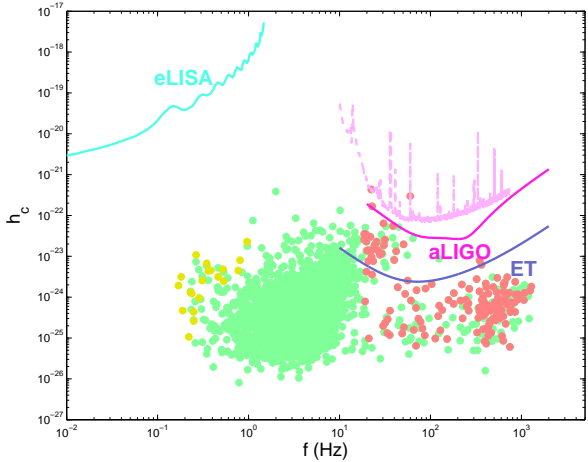}
\caption{Estimations of the characteristic gravitational wave amplitude $h_c$ for Galactic pulsars and magnetars, compared against the noise curves of various GW detectors, i.e. aLIGO (pink, O1 [dashed] and the full design [solid] sensitivity), ET (blue) and eLISA (cyan). The orange points present the selected pulsars in the aLIGO O1 results, while the green points are other Galactic pulsars.} 
\label{fig:2}
\end{figure}

We plot the estimated $h_c$ values (with $\eta_m/\eta_p=1$) for all the pulsars and magnetars in Figure \ref{fig:2}. This is compared against the sensitivities of the GW detectors, i.e. aLIGO, ET and eLISA, for a single detector analysis. 
We find that for $\eta_m/\eta_p = 1$, the GW signals from these pulsars are not detectable for the aLIGO detector at the full design. The eLISA detector, limited by its designed sensitivity, is also not suitable for detecting the signals from Galactic magnetars or known pulsars. The ET detector may be able to detect some relatively low frequency signals ($<50$ Hz) from some of these pulsars. It is worth noticing that although the magnetic field strength of the Galactic magnetars are similar to millisecond magnetars, their characteristic GW amplitudes are quite low due to their much slower spin period. 

In the above analysis we only compare the $h_c$ defined in Eq. \ref{eq:hc}  and the detector sensitivity defined in Eq. \ref{eq:hrms} to estimate the detectability of the GW signal. It is possible to implement a more comprehensive coherent data analysis procedure \citep{cutler05,dupuis05,astone10} to improve the GW signal detection \citep{aasi15}. With such a technique, a few pulsars shown in Figure \ref{fig:2} may become detectable by aLIGO. Considering that the LIGO detectors are escalating and the ET detectors are still in the stage of conceptual development, we would like to leave a detailed investigation of such an effect to future work.

\section{Conclusion and Discussion}

Rapidly rotating neutron stars are potential sources of continuous gravitational waves for ground-based interferometric GW detectors, if the neutron stars may sustain a non-axisymmetric deformation. Recently, $\epsilon \sim 0.005$ for rapidly spinning, strongly magnetized, supramassive neutron stars (millisecond magnetars) have been inferred from the statistical observational properties of Swift SGRBs.  We estimate the detection horizon of such millisecond magnetars by the current (aLIGO) and future (ET) GW detectors. For fast rotators ($P \sim 1$ ms), the horizon is $\sim 30$ Mpc and $\sim 600$ Mpc, respectively, for aLIGO and ET. For slow rotators (e.g. $P \sim 8$ ms), the horizon can be extended to $\sim 400$ Mpc and $\sim 3$ Gpc, respectively. The non-detection of such millisecond magnetars from SGRBs by aLIGO is consistent with the inferred short period ($\sim 1$ ms) of these magnetars \citep{gao16}.

Assuming that magnetic distortion is the main origin of ellipticity, in this work we show that these values are consistent with the non-detection results of Galactic pulsars by aLIGO O1, as long as $\eta_m/\eta_p$ is not greater than 1 by more than one order of magnitude. 
We further estimate the characteristic gravitational wave amplitude $h_c$ for known pulsars and normal magnetars and find that the GW signals from these pulsars are not detectable by the aLIGO detector full design and by eLISA (assuming $\eta_m/\eta_p=1$). The ET detector may be able to detect the relatively low frequency signals ($<50$ Hz) from some of these pulsars.

It is possible that the ellipticity of the millisecond magnetar is not mainly contributed by magnetic deformation. For non-magnetic distortions, the distortion is usually more significant for rapid rotators, so that given the same $\epsilon$ inferred from the millisecond magnetars in SGRBs, the $\epsilon$ for Galactic pulsars/magnetars could be even lower than the $\eta_m/\eta_p=1$ extrapolation shown in Figure \ref{fig:1}. This would be even more consistent with the aLIGO O1 non-detection result, and the detectability of Galactic NSs by future GW detectors would be more pessimistic. 

It is worth pointing out that the SGRB-data-inferred ellipticity value for millisecond magnetars could be inconsistent with the aLIGO O1 results if $\eta_p$ is smaller than $\eta_m$ by more than one order of magnitude. However, according to previous studies \citep{bonazzola96b,konno00,stella05,haskell08,mastrano11}, in order to achieve $\epsilon\sim 0.005$, a very high strength ($10^{16-17}$ G) is needed, implying that the internal (toroidal) field of the millisecond magnetar may be more than $1-2$ orders of magnitude stronger than the dipole field value $10^{15}$ G, namely $\eta_m\sim 0.01-0.1$. In this case, the toroidal field of the Galactic pulsars need to be more than 3 orders of magnitude stronger than the dipole field value ($\eta_p$ being smaller than $0.001$) in order to invalidate $\epsilon\sim 0.005$ for millisecond magnetars. This is essentially impossible. There is no evidence of significant toroidal magnetic field component for radio pulsars. Even though it is conjectured that a toroidal component exists for Galactic magnetars, the degree is twisting must be much weaker than millisecond magnetars since  the magnetar activities (quiescent emission and soft $\gamma$-ray bursts) are believed to be powered by magnetar untwisting \citep{thompson01}.
Very likely $\eta_m/\eta_p$ is less than unity instead, so that the characteristic gravitational wave amplitude $h_c$ for Galactic pulsars and magnetars shown in Figure \ref{fig:2} are over estimated. Even ET might not be capable to detect these sources. 

When estimating the detection probability of Galactic pulsars and magnetars for aLIGO full design, ET and eLISA, we simply compare the characteristic gravitational wave amplitude $h_c$ of the sources with the analytical noise curve of the detectors. In reality, the noise curves may become more complicated due to some additional noises (see aLIGO O1 curve above the analytical aLIGO full-design curve in Figure \ref{fig:2}). This would drop the signals that are only slightly above the noise curve, rendering them not detectable (see \citep{abbott17} for examples). 
On the other hand, a more comprehensive coherent data analysis procedure would improve the GW signal detection probability. It is possible that a few GWs from the pulsars shown in Figure \ref{fig:2} may become detectable even by aLIGO with the help of such a technique.

We thank the referee for the helpful comments which have helped us to improve the presentation
of the paper. This work is supported by the National Basic Research Program (`973' Program) of China (grants 2014CB845800), the National Natural Science Foundation of China under grants 11690024, 11603003, 11633001. Z Cao was supported by ``the Fundamental Research Funds for the Central Universities".

{}

\end{document}